\documentclass[11pt]{article}
\newcommand{\barre}[1]{%
	\setbox1=\hbox{$#1$} \dimen2=\ht1 \dimen3=\dp1 \dimen4=\wd1
	\setbox2=\hbox{\sl /}
	\dimen1=\wd1 \advance\dimen1 by -\wd2 \divide\dimen1 by 2
	\advance\dimen1 by \wd2 \advance\dimen1 by 0.4pt
	\setbox3=\hbox to \wd1{\hss \box1 \kern -\dimen1 \box2\hss}
	\ht3=\dimen2 \dp3=\dimen3 \wd3=\dimen4
	\box3
	}

\newcommand{\vev}[1]{%
	\langle #1 \rangle
	}

\newcommand{\svev}[1]{%
	|\langle #1 \rangle|^2
	}

\begin{document}
\newcommand{\be}{\begin{equation}}
\newcommand{\ee}{\end{equation}} 
\newcommand{\bea}{\begin{eqnarray}}
\newcommand{\eea}{\end{eqnarray}} 
\newcommand{\nn}{\nonumber}
\newcommand{\bal}{\begin{array}{ll}} 
\newcommand{\eal}{\end{array}}
\newcommand{\la}[1]{\lambda^{#1}}
\newcommand{\Tr}{\rm Tr}
\def\1{{\rm 1 \kern -.10cm I \kern .14cm}} \def\R{{\rm R \kern -.28cm I
\kern .19cm}}


\begin{titlepage}
\begin{flushright}   UFIFT-HEP-97-34 \\ hep-ph/9712239 \end{flushright}
\vskip 2.5cm
\centerline{\LARGE{\bf {A guide to flat direction analysis}}}
\vskip .5cm
\centerline{\LARGE{\bf {in anomalous $U(1)$ models}}}
\vskip 1cm
\centerline{\bf Nikolaos Irges${}^{\,}$\footnote{Supported in part by the
United States Department of Energy under grant DE-FG05-86-ER40272.}${}^{,}$\footnote{E-mail address: irges@phys.ufl.edu.} and St\'ephane Lavignac${}^{\,}$\footnote{Attach\'e Temporaire d'Enseignement et de Recherche at Universit\'e Paris VII, Paris, France.}${}^{,}$\footnote{Permanent address: Laboratoire de Physique Th\'eorique et Hautes
Energies, Universit\'e Paris-Sud, B\^at. 210, F-91405 Orsay Cedex, France (e-mail address: lavignac@qcd.th.u-psud.fr).}}
\vskip .5cm
\centerline{\em Institute for Fundamental Theory,}
\centerline{\em Department of Physics, University of Florida}
\centerline{\em Gainesville FL 32611, USA}

\vskip1.5cm
\centerline{\bf {Abstract}}
\indent

We suggest a systematic procedure to study $D$- and $F$-flat directions in a large class of models with an anomalous $U(1)$. This class of models is characterized by the existence of a vacuum that breaks all Abelian gauge symmetries connecting the observable sector to the hidden sector. We show that, under some conditions, there is no other stable vacuum that breaks these symmetries. As a consequence, the model yields definite (order of magnitude) predictions for low-energy mass hierarchies. Then we study generic flat directions and identify the ones that may lead to undesirable vacua. We give necessary conditions for those to be lifted, and show that supersymmetry breaking only slightly affects the conclusions from the flat direction analysis.

\vfill
\begin{flushleft}
December 1997 \\
\end{flushleft}
\end{titlepage}

\section{Introduction}
\label{section:introduction}
\indent

There has been a growing interest for models with an anomalous Abelian gauge symmetry in the last few years. This anomalous $U(1)$, which is a generic prediction of a wide class of effective string theories \cite{string_models}, has been shown to play a role in several fundamental issues such as the origin of fermion mass hierarchies \cite{fermion_masses}, supersymmetry breaking \cite{susy_breaking}, $R$-parity conservation \cite{BILR} and inflation \cite{inflation}. Beyond the fact that it has non-vanishing anomalies, to be compensated for by the Green-Schwarz mechanism \cite{Green-Schwarz}, the most remarkable feature of an anomalous $U(1)_X$ is the structure of its $D$-term:
\begin{equation}
  D_X\ =\ \sum_i X_{\Phi_i}\, |\Phi_i|^2\ -\ \xi^2
\label{eq:D_X}
\end{equation}
where the Fayet-Iliopoulos term $\xi^2$ is computed to be\footnote{In our conventions, the $X$ charge is normalized in such a way that $C_g = \Tr\, X$ is negative.} $(- \Tr\, X)\, g^2\, M_{Pl}^2 / 192\, \pi^2$ in string theory \cite{Atick}. For supersymmetry to be preserved, some field must acquire a vacuum expectation value (vev) of order $\xi$. As a result, $U(1)_X$ is broken slightly below the Planck scale \cite{DSW}, and its breaking scale is fixed by the mixed gravitational anomaly $C_g = \Tr\, X$.

The vacuum structure of models with an anomalous $U(1)$ is therefore very particular, and the issue of characterizing the flat directions along which it is broken is a very crucial one\footnote{The first study of flat directions in a model with an anomalous $U(1)$ and a singlet field beyond the MSSM has been done in Ref. \cite{DGPS}.}. First because these flat directions may lead to non-viable vacua in which either supersymmetry (if (\ref{eq:D_X}) has no solution, or if its solutions are spoiled by $F$-terms) or the Standard Model symmetries (if e.g. a squark acquires a vev of order $\xi$) are broken at the scale $\xi$ \footnote{As was shown in Ref. \cite{BILR}, the requirement that this does not happen puts severe constraints on the models, which in turn provides valuable information on issues such as $R$-parity and the origin of the mu-term.}. Secondly, the phenomenological implications of the anomalous $U(1)$ strongly depend on the nature of the low-energy vacuum. For example, in fermion mass models based on horizontal Abelian symmetries \cite{Froggatt-Nielsen}, the Yukawa couplings are functions of singlet vevs which couple to the quarks and leptons. A similar statement can be made for neutrinos masses and mixings \cite{neutrinos}, sfermion masses and the magnitude of flavour changing neutral currents \cite{FCNC,DGPS}, as well as baryon and lepton number violating couplings \cite{B_and_L}.

In this letter, we suggest a systematic step by step procedure to analyze $D$-flat and $F$-flat directions in such models. In section \ref{section:FD}, we first define the class of anomalous $U(1)$ models to which our analysis apply. Then we solve the $D$-term constraints, and we study under which conditions the $D$-flat directions are lifted by $F$-terms. In section \ref{section:susy}, we show the relevance of the flat direction analysis to the determination of the low-energy vacuum. The procedure is summarized in Section \ref{section:procedure}, and it is applied to an explicit model in Section \ref{section:model}. We state our conclusions in Section \ref{section:conclusion}.

\section{Flat direction analysis}
\label{section:FD}

\subsection{Models with an anomalous $U(1)$}
\label{subsection:assumptions}
\indent

We consider supersymmetric models with a gauge group $G_{obs} \times U(1)_X \times U(1)_1 \times \ldots \times U(1)_N \times G_{hid}$, where $G_{obs}$ contains either the Standard Model or a GUT group, $G_{hid}$ is some hidden gauge group, and there is a set of Abelian (horizontal) factors connecting both sectors. We denote the anomalous charge by $X$, and the non-anomalous ones by $Y_1$, \ldots, $Y_N$. In general, but not always, fields charged under $G_{obs}$ are singlets under $G_{hid}$ and vice versa\footnote{This particularity enables us to study separately the flat directions involving observable and hidden fields. Our main interest in this paper is the observable sector, but our analysis applies to the hidden sector as well.}, both carrying $X$, $Y_1$, \ldots and $Y_N$ charges. This structure is characteristic of several classes of string models \cite{Faraggi,Kim,Chaudhuri,Tye}. In the following, we call $G = G_{obs} \times G_{hid}$, and we denote generically the fields charged under $G$ by $\Phi_i\,$, and the $G$-singlets by $\chi_i\,$.

As stressed in the introduction, some of these fields must acquire non-vanishing vacuum expectation values through the Dine-Seiberg-Witten (DSW) mechanism \cite{DSW} in order for supersymmetry to be preserved. This in turn breaks $U(1)_X$ slightly below the string scale, possibly together with some other symmetries. Since the Standard Model symmetries must not be broken at that scale, we shall assume that there exists a solution of the $D$- and $F$-term equations that breaks only the Abelian factors\footnote{This assumption may be not very convenient for models that have a unified gauge group in the observable sector, because this group must break down at the $GUT$ scale. However, in the models of Ref. \cite{Tye}, the adjoint Higgs fields needed to break the GUT group do not carry any Abelian charges, so that they can develop vevs without any effect on the Abelian $D$-terms and their solutions.}. The $D$-terms equations:
\begin {equation}
  \begin{array}{lllllll}
  D_X & = & \sum_{\alpha} x_{\alpha}\, \svev{\theta_{\alpha}} & - & \xi^2 & = & 0  \\   \\
  D_{Y_i} & = & \sum_{\alpha}\ y^{[i]}_{\alpha}\, \svev{\theta_{\alpha}} & & & = & 0
  \end{array}
\label{eq:DSW_eq}
\end{equation}
where the $G$-singlets with non-vanishing vevs are denoted by $\theta_{\alpha}$, have in general several solutions, due to the large number of $G$-singlets generally present in  string models. We shall assume the existence of at least one solution \{$\vev{\theta_{\alpha}}$\} of (\ref{eq:DSW_eq}) satisfying the following requirements:
\begin{itemize}
  \item  all Abelian symmetries connecting the hidden sector to the observable sector are broken at the scale $\xi$; while probably too strong, this requirement enables the models to escape many phenomenological problems\footnote{Note that the Abelian factors that $G$ may contain, being either observable or hidden, do not suffer from these problems. So they can survive down to low energies.}. The number of $\theta$ fields must then be equal to the number of $U(1)$'s or greater.
  \item the low-energy mass hierarchies (in particular fermion masses), which are generated by the small parameters $\vev{\theta_{\alpha}} / M_{Pl}$, are {\it completely determined} by the high-energy theory. This means that there must be no more $\theta$ fields than $U(1)$'s, otherwise the $\vev{\theta_{\alpha}}$ would not be uniquely determined by (\ref{eq:DSW_eq}).
\end{itemize}

In other words, we assume the existence of at least one ($N+1$)-plet of $G$-singlets ($\theta_0$, \ldots, $\theta_N$) such that:

{\bf (a)} the matrix of the $\theta$ field charges is invertible:
\begin{equation}
  \det A \neq 0  \hskip .8cm  \mbox{where}  \hskip .8cm
	A\ =\ \left( \begin{array}{llll}
  x_0 & x_1 & \ldots & x_N  \\
  y^{[1]}_0 & y^{[1]}_1 & \ldots & y^{[1]}_N  \\
  \ldots & \ldots & \ldots & \ldots \\
  y^{[N]}_0 & y^{[N]}_1 & \ldots & y^{[N]}_N
	\end{array} \right)
\label{eq:condition1}
\end{equation}
and the first column of $A^{-1}$ only contains strictly positive entries. This ensures the existence of a vacuum $\vev{\theta_0, \ldots, \theta_N}_{DSW}$ with
\begin{equation}
  \left( \begin{array}{c} \svev{\theta_0}_{DSW} \\ \svev{\theta_1}_{DSW}  \\ \vdots \\ \svev{\theta_N}_{DSW} \end{array} \right)\ =\ A^{-1}\ \left( \begin{array}{c} \xi^2 \\ 0 \\ \vdots \\ 0 \end{array} \right)
\label{eq:DSW}
\end{equation}
which we shall refer to as the {\it DSW vacuum}.

In addition, one must check that this vacuum is not spoiled by the $F$-terms. Since condition (\ref{eq:condition1}) ensures that there is no invariant of the form $\theta^{n_0}_0 \theta^{n_1}_1 \ldots \theta^{n_N}_N$, this can happen only if the superpotential contains a term linear in the $\chi$ and $\Phi$ fields. Our second assumption is then:

{\bf (b)} there is no holomorphic invariant $\chi\, \theta^{n_0}_0 \theta^{n_1}_1 \ldots \theta^{n_N}_N$ linear in $\chi$, where $\chi$ is a $G$-singlet other than the $\theta$ fields. This amounts to a condition on the charges of $\chi$, namely at least one of the numbers $n_0$, \ldots, $n_N$ defined by
\begin{equation}
  \left( \begin{array}{c} n_0 \\ n_1  \\ \vdots \\ n_N \end{array} \right)\ =\ -\ A^{-1}\ \left( \begin{array}{c} X_{\chi} \\ Y^{[1]}_{\chi}  \\ \vdots \\ Y^{[N]}_{\chi} \end{array} \right)
\label{eq:condition2}
\end{equation}
has to be either fractional or negative. Note that if $\chi$ is a right-handed neutrino, this constraint leads to an automatic conservation of $R$-parity \cite{BILR}. Finally, let us stress that the additional constraint $\vev{W}=0$, which must be imposed in (unbroken) local supersymmetry, is automatically satisfied by the DSW vacuum: in the absence of an invariant $\theta^{n_0}_0 \theta^{n_1}_1 \ldots \theta^{n_N}_N$, any term in $W$ gives a vanishing contribution to $\vev{W}$.

Since we are dealing with an effective field theory, we must put in the superpotential all possible interactions allowed by the symmetries of the theory, including non-renormalizable terms suppressed by inverse powers of the Planck mass (in the following, we set $M_{Pl}=1$). An important comment is in order here. Due to discrete symmetries and conformal selection rules, the superpotential of an effective string theory does not contain every term allowed by the (continuous) gauge symmetries. This may have important consequences, in particular some $D$-flat directions that one would naively expect to be lifted by $F$-terms could remain flat to all orders \cite{Cleaver}. In order to keep our discussion as general as possible, {\it we shall neglect this effect for the rest of the paper}. Thus the criteria that we give in Subsection \ref{subsection:F_terms} for a $D$-flat direction to be lifted should be regarded as necessary conditions only.

Let us have a look at the generic form of superpotential terms. In general, $G$-invariants and $G$-singlets are not neutral under the Abelian symmetries, and must appear multiplied by powers of the $\theta$ fields. Condition (a) allows us to assign, through Eq. (\ref{eq:condition2}), a set of numbers \{$n_{\alpha}$\} $\equiv$ ($n_0$, \ldots, $n_N$) to each $G$-invariant $S = \Phi^{p_1}_1 \ldots \Phi^{p_n}_n$ (resp. $G$-singlet $\chi$). If all $n_{\alpha}$ are positive integers, then $S\, \theta^{n_0}_0 \theta^{n_1}_1 \ldots \theta^{n_N}_N$ is an holomorphic invariant and can be present in the superpotential\footnote{It is quite remarkable that condition (\ref{eq:condition1}), which ensures the existence of the DSW vacuum, is at the same time the one that is required for invariants of the form $S\, \theta^{n_0}_0 \theta^{n_1}_1 \ldots \theta^{n_N}_N$ to exist \cite{EIR}.  Those invariants are precisely the ones needed to generate mass hierarchies in the DSW vacuum, with $S$ being Yukawa couplings. Note that this is not true in the non-anomalous case: the condition required for the $\theta$ fields to develop a nonzero vev is $\det A = 0$, which forbids the existence of the invariants $S\, \theta^{n_0}_0 \theta^{n_1}_1 \ldots \theta^{n_N}_N$, except for some very specific charge assignments for which the powers ($n_0$, \ldots, $n_N$) are not uniquely determined.}. If all $n_{\alpha}$ are positive, but some of them are fractional, the invariant appears at higher order: $\left( S\, \theta^{n_0}_0 \theta^{n_1}_1 \ldots \theta^{n_N}_N \right)^m$. Finally, if some $n_{\alpha}$ is negative, one can not form any holomorphic invariant out of $S$ and the $\theta$ fields; we shall refer to this last situation by saying that $S$ corresponds to a {\it supersymmetric zero} in the superpotential.

\subsection{$\mbox{\boldmath $D$}$-flatness}
\label{subsection:D_terms}
\indent

Before characterizing the $D$-flat directions of the models defined above, let us recall a very usefull theorem \cite{Buccella} which we shall use throughout this paper. In a globally supersymmetric theory with a compact gauge group $G$, and {\it no} Fayet-Iliopoulos term associated with the Abelian factors that $G$ may contain, the zeroes of the $D$-terms can be classified in terms of the holomorphic gauge invariants. More precisely, a set of vevs $\vev{\Phi_1, \ldots, \Phi_n}$ is a solution of the $D$-term constraints if and only \cite{Gatto} if there exists a $G$-invariant holomorphic polynomial $I(\Phi_1, \ldots, \Phi_n)$ such that:
\begin{equation}
  \left. {\partial{\, I} \over \partial{\Phi_i}}\, \right|_{\Phi_i=\vev{\Phi_i}}\! =\ C\, \vev{\Phi^{\dagger}_i}\  \hskip 2cm i\ =\ 1 \ldots n
\label{eq:ansatz}
\end{equation}
where $C$ is a complex dimensional constant. A systematic way to study $D$-flat directions is then to find a finite basis of invariant monomials \{$S_a$\} over which any holomorphic invariant polynomial can be decomposed. Such a basis is characteristic of the gauge group and the field content of the theory. As an example, a basis of the MSSM invariants can be found in Ref. \cite{Martin}.

We are now ready to make the following statement: {\it to each basis $G$-invariant $S = \Phi^{p_1}_1 \ldots \Phi^{p_n}_n$ (resp. $G$-singlet $\chi$), corresponds a $D$-flat direction $\vev{\Phi_1, \ldots \Phi_n\,; \vec{\theta}\,}$ (resp. $\vev{\chi\,; \vec{\theta}\,}$), with}
\begin{equation}
  \left\{  \begin{array}{lll}
  \svev{\theta_{\alpha}}\ \geq\ \svev{\theta_{\alpha}}_{DSW} & \hskip 1cm & \mbox{for} \hskip .3cm n_{\alpha}\ \geq\ 0  \\
    &  &  \\
  \svev{\theta_{\alpha}}\ <\ \svev{\theta_{\alpha}}_{DSW} & \hskip 1cm & \mbox{for} \hskip .3cm n_{\alpha}\ <\ 0
  \end{array}  \right.
\label{eq:theta_constraints}
\end{equation}
As we show below, {\it this is the only solution of the $D$-term constraints associated with $S$ (resp. $\chi$)}. Note that this is not, in general, a flat direction of the scalar potential, because the $F$-term constraints $F_i = 0$ are not necessarily satisfied.

Let us prove this first in the case of a $G$-singlet  $\chi$. The only input we need is the existence of the DSW vacuum (\ref{eq:DSW}). Then, using the definition (\ref{eq:condition2}) of the \{$n_{\alpha}$\}, the $D$-term constraints
\begin{equation}
  \left( \begin{array}{c} X_{\chi} \\ Y^{[1]}_{\chi}  \\ \vdots \\ Y^{[N]}_{\chi} \end{array} \right)\ \svev \chi\ +\ A\ \left( \begin{array}{c} \svev{\theta_0} \\ \svev{\theta_1}  \\ \vdots \\ \svev{\theta_N} \end{array} \right)\ =\ \left( \begin{array}{c} \xi^2 \\ 0 \\ \vdots \\ 0 \end{array} \right)
\label{eq:Dterms}
\end{equation}
can be rewritten as:
\begin{equation}
  \left( \begin{array}{c} \Delta\, \svev{\theta_0} \\ \Delta\, \svev{\theta_1}  \\ \vdots \\ \Delta\, \svev{\theta_N} \end{array} \right)\ =\ \svev \chi\ \left( \begin{array}{c} n_0 \\ n_1  \\ \vdots \\ n_N \end{array} \right)
\label{eq:Dflat}
\end{equation}
where we have defined $\Delta\, \svev{\theta_{\alpha}} = \svev{\theta_{\alpha}} - \svev{\theta_{\alpha}}_{DSW}$. The sign of $\Delta\, \svev{\theta_{\alpha}}$ is thus determined by the sign of $n_{\alpha}$. In particular, when the $D$-flat direction is associated with an holomorphic invariant of the whole gauge group $G \times U(1)_X \times U(1)_1 \times \ldots \times U(1)_N$, constraints (\ref{eq:theta_constraints}) read:
\begin{equation}
  \svev{\theta_{\alpha}}\ \geq\ \svev{\theta_{\alpha}}_{DSW}  \hskip 2cm  \alpha\ =\ 0, 1 \ldots N
\label{eq:theta_holomorphic}
\end{equation}
This is a remarkable difference with the non-anomalous case, in which the vevs of the $\theta$ fields are not bounded. The $\Delta\, \svev{\theta_{\alpha}}$ depend on a single parameter\footnote{This is no longer true when $\chi$ is a singlet of the whole gauge group. In this case, $n_0 = n_1 = \ldots = n_N = 0$ and (\ref{eq:Dflat}) is nothing but the DSW vacuum, whatever $\vev{\chi}$ may be.}, $\svev \chi$, which may be fixed by the $F$-term constraints, or by supersymmetry breaking. In the particular case where $\vev \chi = 0$, one recovers the DSW vacuum.

The generalization of (\ref{eq:Dflat}) to the case of a basis $G$-invariant is straightforward. Applying (\ref{eq:ansatz}) to $S = \Phi^{p_1}_1 \ldots \Phi^{p_n}_n$, we find that the $D$-terms associated with $G$ constrain the vevs of the $\Phi$ fields to be aligned\footnote{In this relation, $\svev{\Phi_i}$ stands for $\sum_{\alpha} \svev{\Phi^{\alpha}_i}$, where the $\Phi^{\alpha}_i$ are the components of the representation of $G$ spanned by $\Phi_i$. One should keep in mind that (\ref{eq:alignment}) is a weaker constraint than the vanishing of the $D$-terms.}:
\begin{equation}
  {\svev{\Phi_1} \over p_1}\ =\ \ldots\ =\ {\svev{\Phi_n} \over p_n}\ \equiv\ v^2_S
\label{eq:alignment}
\end{equation}
As a result, we end up with a relation similar to (\ref{eq:Dflat}), with $\svev \chi$ replaced by $v^2_S$. In the following, we shall denote this $D$-flat direction by $\vev{S, \vec{\theta}\,}$ to stress the fact that the vevs of the fields in $S$ are aligned.

However, generic $D$-flat directions are not associated with a single basis $G$-invariant, but rather with a polynomial in the basis $G$-invariants. More precisely, flat directions involving a given set \{$\Phi_i$\} of fields charged under $G$ are parameterized by the vevs of the $G$-invariants $S_a = \prod_i \Phi^{p^a_i}_i$ that can be formed out of those fields\footnote{Since the basis invariants are in general not independent, this parameterization is obviously redundant, but it turns out to be very convenient for our purpose.}:
\begin{equation}
  \svev{\Phi_i}\ =\ \sum_a\ v^2_a\ p^a_i
\label{eq:alignment2}
\end{equation}
where $v^2_a$ is a vev\footnote{In general, the parameters $v^2_a$ are complex. In the following, we shall assume that they can always be chosen to be positive real numbers. While we do not have a general proof for this, it turns out to be the case in numerous explicit examples. We thank Carlos Savoy for a discussion on this point.} associated with the invariant $S_a$. Then the most general solution of the complete set of $D$-term constraints is a set of vevs $\vev{\{\Phi_i\}, \{\chi_i\}, \vec{\theta}\,}$ with:
\begin{equation}
  \left( \begin{array}{c} \Delta\, \svev{\theta_0} \\ \Delta\, \svev{\theta_1}  \\ \vdots \\ \Delta\, \svev{\theta_N} \end{array} \right)\ =\ \sum_a\ v^2_a\ \left( \begin{array}{c} n^a_0 \\ n^a_1  \\ \vdots \\ n^a_N \end{array} \right)\ +\ \sum_i\ \svev{\chi_i}\ \left( \begin{array}{c} n^i_0 \\ n^i_1  \\ \vdots \\ n^i_N \end{array} \right)
\label{eq:Dflat2}
\end{equation}
Clearly the relation $\svev{\theta_{\alpha}} \geq \svev{\theta_{\alpha}}_{DSW}$ holds, for a given $\alpha$, only when all powers $n^a_{\alpha}$ and $n^i_{\alpha}$ are positive. On the contrary, when one of these numbers is negative, $\svev{\theta_{\alpha}}$ can be smaller than $\svev{\theta_{\alpha}}_{DSW}$. This may lead to vacua in which $\vev{\theta_{\alpha}}$ vanishes after imposing the $F$-term constraints, or after supersymmetry breaking. We shall see in the following that formulae (\ref{eq:Dflat}) and (\ref{eq:Dflat2}) considerably simplify the analysis of flat directions in anomalous $U(1)$ models.

\subsection{$\mbox{\boldmath $F$}$-flatness}
\label{subsection:F_terms}
\indent

In this section, we examine under which conditions a $D$-flat direction is lifted by $F$-terms. We first assume no compensation between different contributions to the $F$-terms, so that each individual contribution has to vanish for a $D$-flat direction to be preserved. We shall come back to this point in Subsection \ref{subsubsection:cancellations}.

\subsubsection{Flat directions associated with a single $G$-invariant}
\label{subsubsection:single_S}
\indent

We first restrict our attention to $D$-flat directions $\vev{S, \vec{\theta}\,}$ that are associated with a single $G$-invariant $S = \Phi^{p_1}_1 \ldots \Phi^{p_n}_n$. Two cases must be distinguished, depending on the signs of the numbers $n^S_{\alpha} = (n^S_0, \ldots, n^S_N)$ associated with $S$:
\begin{itemize}
  \item {\bf all $n^S_{\alpha}$ are positive}, i.e. the $D$-flat direction can be associated with some holomorphic invariant $S^m\, \theta^{m_0}_0 \theta^{m_1}_1 \ldots \theta^{m_N}_N$ of the whole gauge group. This invariant contributes to the $F$-terms as\footnote{Let us recall here that throughout this paper, we assume that the superpotential contains every term allowed by (continuous) gauge invariance and holomorphy. This is not the case in string models, where discrete symmetries and selection rules generally forbid a lot of terms.} 
\begin{equation}
  \vev{F_{\Phi_i}}\ =\ \alpha\, m\, C^m\, v_S^{2(m-1)}\ \vev{\Phi^{\dagger}_i}\ \vev{\theta_0}^{m_0}\, \vev{\theta_1}^{m_1} \ldots\, \vev{\theta_N}^{m_N}\ +\ \ldots
\end{equation}
where we made use of (\ref{eq:ansatz}), $\alpha$ is a coupling constant, and $v_S$ is defined by (\ref{eq:alignment}). Since $\svev{\theta_{\alpha}} \geq \svev{\theta_{\alpha}}_{DSW}$ along the flat direction, this contribution vanishes only if $\vev{\Phi_i}=0$, or equivalently $v_S=0$. As a result, the flat direction breaks down to the DSW vacuum.
  \item {\bf some of the $n^S_{\alpha}$ are negative}, i.e. the $D$-flat direction cannot be associated with  any holomorphic invariant of the whole gauge group. Such flat directions are in general not completely lifted, unless the superpotential contains an invariant of the form $S'\, \theta^{n'_0}_0 \theta^{n'_1}_1 \ldots \theta^{n'_N}_N$ (with $S'$ a combination of basis $G$-invariants and $\chi$ fields), where either one of the following two conditions is fulfilled\footnote{When several $n_{\alpha}$ are negative, these conditions are actually too strong. Using (\ref{eq:Dflat}), one can show that, in most cases, only one $n'_{\alpha}$ has to vanish.}: (i) $S'$ contains no other field than the ones appearing in $S$, and $n'_{\alpha}=0$ or $1$ if $n_{\alpha}<0$ (with the additional constraint $\sum_{\, \{\alpha;\, n_{\alpha}<0\}}\, n'_{\alpha} \leq 1$); (ii) $S'$ contains only one field that does not appear in $S$, and $n'_{\alpha}=0$ if $n_{\alpha}<0$.
\end{itemize}

To illustrate the last point, let us consider a toy model with $G = SU(3)_C \times SU(2)_L \times U(1)_Y$, two $U(1)$'s, and the following field content: $Q_1$, $\bar u_1$, $\bar u_2$, $H_u$. We assume that the invariant associated with $S \equiv Q_1 \bar u_1 H_u$ is non-holomorphic, e.g. $Q_1 \bar u_1 H_u\, \theta^{-1}_0 \theta^{3}_1$. Then the superpotential contains only one term:
\begin{equation}
  W\ =\ \alpha\ Q_1 \bar u_2 H_u\, \theta^{n_0}_0 \theta^{n_1}_1
\end{equation}
provided that $n_0$ and $n_1$ are positive integers. Note that $S' \equiv Q_1 \bar u_2 H_u$ contains only one field, $\bar u_2$, that does not appear in $S$; therefore, it satisfies condition (ii) provided that $n_0=0$. The only $F$-terms that are likely to be non-vanishing along the $D$-flat direction $\vev{Q_1, \bar u_1, H_u, \theta_0, \theta_1}$ associated with $S$ are:
\begin{eqnarray}
  F_{\bar u_2}\ =\ \alpha\ Q_1 H_u\, \theta^{n_0}_0 \theta^{n_1}_1
\label{eq:F_u2}
\end{eqnarray}
 Using (\ref{eq:ansatz}), we obtain:
\begin{equation}
  \vev{F_{\bar u_2}}\ =\ \alpha\, C\, \vev{\bar u^{\dagger}_1}\ \vev{\theta_0}^{n_0} \vev{\theta_1}^{n_1}
\label{eq:contribution1}
\end{equation}
If $n_0=0$, the $F$-terms vanish only for $v_S=0$, and the $D$-flat direction $\vev{Q_1, \bar u_1, H_u, \theta_0, \theta_1}$ breaks down to the DSW vacuum $\vev{\theta_0, \theta_1}_{DSW}$. If $n_0>0$, the $F$-term constraints have two solutions: the DSW vacuum (which corresponds to $v_S=0$), and a residual flat direction $\vev{Q_1, \bar u_1, H_u, \theta_1}$ with $v^2_S \equiv \svev{Q_1} = \svev{\bar u_1} = \svev{H_u} = \svev{\theta_0}_{DSW}$ and $\svev{\theta_1} = 3\, \svev{\theta_0}_{DSW} + \svev{\theta_1}_{DSW}$. Thus the initial $D$-flat direction is only partially lifted, and the residual flat direction, along which $\vev{\theta_0} = 0$, can lead to another vacuum than the DSW vacuum. However, it can still be lifted by higher order operators. Indeed, the invariant $\left( Q_1 \bar u_1 H_u \right)^{n_0} \left( Q_1 \bar u_2 H_u \right) \theta^{3n_0+n_1}_1$ (which satisfies condition (ii)) contributes to $\vev{F_{\bar u_2}}$ as:
\begin{equation}
  \beta\, C^{n_0+1}\, v^{2n_0}_S\ \vev{\bar u^{\dagger}_1}\ \vev{\theta_1}^{3n_0+n_1}
\label{eq:contribution2} 
\end{equation}
which obviously vanishes only in the DSW vacuum ($v_S=0$).

\subsubsection{Flat directions of $G$-singlets}
\label{subsubsection:singlets}
\indent

We consider now all possible $D$-flat directions involving only $G$-singlets. Two cases must be distinguished, depending on the signs of the numbers $n^i_{\alpha} = (n^i_0, \ldots, n^i_N)$ associated with each of the fields $\chi_i$:

\begin{itemize}
  \item  {\bf all $n^i_{\alpha}$ are positive:} in this case, $\svev{\theta_{\alpha}} \geq \svev{\theta_{\alpha}}_{DSW}$ along any flat direction of singlets. As a consequence, the $D$-term equations do not allow for any other vacuum of singlets than the vacuum (\ref{eq:DSW}). The other solutions of (\ref{eq:DSW_eq}) are $D$-flat directions parameterized by the vevs of the $\chi$ fields. Since these flat directions correspond to holomorphic invariants of the whole gauge group, they are lifted by $F$-terms, leaving only\footnote{While compensations inside the $F$-terms may lead to other vacua than the vacuum (\ref{eq:DSW}), we shall consider this possibility as very unlikely (see Subsection \ref{subsubsection:cancellations}).} the vacuum (\ref{eq:DSW}). Therefore, in this case, {\it the DSW vacuum is unique}.
  \item  {\bf some $n^i_{\alpha}$ are negative:} in this case, some of the $\svev{\theta_{\alpha}}$ can be smaller than in vacuum (\ref{eq:DSW}). As can be seen from (\ref{eq:Dflat2}), the $D$-term equations allow for vacua of singlets in which $\vev{\theta_{\alpha}} = 0$, while some of the $\chi$ fields have non-vanishing vevs. Those vacua correspond to particular points along $D$-flat directions that are in general not completely lifted, unless the required holomorphic invariants (see Subsection \ref{subsubsection:single_S}) are present in the superpotential\footnote{See previous note.}. If this is the case, one recovers the uniqueness of the DSW vacuum.
\end{itemize}

\subsubsection{Generic flat directions}
\label{subsubsection:generic}
\indent

Consider now a generic flat direction $\vev{\{\Phi_i\}; \{\chi_i\}; \vec{\theta}\,}$ involving $G$-charged fields as well as $G$-singlets. The relevant numbers here are \{$n^a_{\alpha}\,$; $n^i_{\alpha}$\}, where the \{$n^a_{\alpha}$\} are associated with the basis $G$-invariants \{$S_a$\} that contain the \{$\Phi_i$\}. The general requirement for this flat direction to be lifted is that invariants of the form $S'\, \theta^{n'_0}_0 \theta^{n'_1}_1 \ldots \theta^{n'_N}_N$ be present in the superpotential (where $S'$ is a combination of basis $G$-invariants and $\chi$ fields), where either one of the following two conditions is fulfilled: (i) $S'$ contains no other field than the ones appearing in the flat direction, and $n'_{\alpha} = 0$ or $1$ if one of the powers \{$n^a_{\alpha}\,$; $n^i_{\alpha}$\} is negative (with the additional constraint that no more than one such $n'_{\alpha}$ should be equal to $1$); (ii) $S'$ contains only one field that does not appear in the flat direction, and $n'_{\alpha} = 0$ if one of the powers \{$n^a_{\alpha}\,$; $n^i_{\alpha}$\} is negative. Several invariants are in general necessary to lift completely the flat direction.

Clearly those conditions are automatically satisfied when all relevant \{$n^a_{\alpha}\,$; $n^i_{\alpha}$\} are positive. In all other cases, one has to check explicitly that the invariants required are present in the superpotential, even if they appear at high orders.

\subsubsection{Cancellations inside the $\mbox{\boldmath $F$}$-terms}
\label{subsubsection:cancellations}
\indent

So far we did not consider the possibility of compensations between different contributions to the $F$-terms. The effect of such cancellations is to reduce the dimensionality of a $D$-flat direction, while one would naively expect it to be (at least partially) lifted. For instance, in the toy model of Subsection \ref{subsubsection:single_S} (case $n_0>0$), contributions (\ref{eq:contribution1}) and (\ref{eq:contribution2}) cancel against each other in $\vev{F_{\bar u_2}}$ if the following relation between vevs is satisfied:
\begin{equation}
  \alpha\, \vev{\theta_0}^{n_0}\ +\ \beta \left( C\, v^2_S \right)^{n_0} \vev{\theta_1}^{3n_0}\ =\ 0
\end{equation}
(note that the case $n_0=0$ does not suffer from this problem, since (\ref{eq:contribution1}) is the only contribution to $\vev{F_{\bar u_2}}$). Such compensations are possible because the $F$-terms, at least at low orders, are not all non-trivial and independent from each other. When higher order operators are added in the superpotential, the number of independent $F$-term constraints generally increases, and cancellations become less likely. We will neglect them here, but in the flat direction analysis of an explicit model, they have to be taken into account.

The case of flat directions involving only $G$-singlets is more subtle, and needs a separate discussion. Due to condition (\ref{eq:condition1}), the $F$-terms of the $\theta$ fields are not independent from the other $F$-terms:
\begin{equation}
  A\ \left( \begin{array}{c} \theta_0\, F_{\theta_0} \\ \vdots \\ \theta_N\, F_{\theta_N} \end{array} \right)\ =\ A_{\chi}\ \left( \begin{array}{c} \chi_1\, F_{\chi_1} \\ \vdots \\ \chi_q\, F_{\chi_q} \end{array} \right)
\label{eq:Fterms}
\end{equation}
where $A_{\chi}$ is the matrix of the charges of the $\chi$ fields, defined in an analogous way to the matrix $A$. As a result, flat directions of $G$-singlets are constrained by exactly as many equations as fields\footnote{This is not the case for flat directions involving fields charged under $G$, because the $D$-terms associated with $G$ provide additional equations.}, and those (non-linear) equations have in general several solutions. Thus the theory possesses, at any order, vacua of singlets that may compete with the DSW vacuum.

However, while the DSW vacuum is well-defined and stable against the addition of higher order terms in the superpotential (as implied by condition (b)), this is obviously not the case for the other vacua of singlets, which depend on the explicit form of the $F$-terms. This would not be a problem if all vevs were small compared with the mass scale by which nonrenormalizable operators are suppressed. But due to the anomalous Fayet-Iliopoulos term, the singlet vevs are generally very close to the Planck scale\footnote{Although we have no general proof for this statement, we could check it in many explicit examples. The reason for this is that in most cases, the other vacua than the DSW vacuum arise due to a series of compensations between terms of various dimensions in the $F$-terms.}, and they are not expected to converge to any fixed value when higher order invariants are added in the superpotential (below we illustrate this point with a simple example). Such a situation obviously does not make sense in the context of an effective field theory, and for this reason we shall consider the DSW solution (in which $\vev{\theta}/M_{Pl}$ is typically of order $0.01 - 0.1$) as the only plausible vacuum of singlets.

To illustrate this point, consider a toy model with three fields $\chi_1$, $\chi_2$ and $\theta$, charged under the gauge group $U(1)_X$ with charges $-5/3$, $-4/3$ and 1. At order 8, the superpotential consists of the following three terms:
\begin{equation}
  W\ =\ c_1 \chi_1^3\theta^5\, +\, c_2 \chi_2^3\theta^4\, +\, c_3 \chi_1\chi_2\theta^3
\label{eq:W_singlets}
\end{equation}
where $c_1$, $c_2$ and $c_3$ are numerical coefficients of order one. If the last term in $W$ were not present, there would be a unique DSW vacuum, with $\svev{\theta}_{DSW} = \xi^2$ and $\vev{\chi_1}_{DSW} = \vev{\chi_2}_{DSW} = 0$. In the presence however of this term, there is an additional solution to the $D$-term and $F$-term constraints: 
\begin{eqnarray}
  \frac{\vev{\chi_1}}{M_{Pl}} & = & \left( -{c_3^3\over 27c_1^2c_2} \right)^{1/3}\ \left( \frac{\vev{\theta}}{M_{Pl}} \right)^{-5/3}  \nonumber  \\
  \frac{\vev{\chi_2}}{M_{Pl}} & = & \left( -{c_3^3\over 27c_1c_2^2} \right)^{1/3}\ \left( \frac{\vev{\theta}}{M_{Pl}} \right)^{-4/3}
\label{eq:vacuum_singlets}
\end{eqnarray}
where due to the positive powers in (\ref{eq:W_singlets}), $|\vev{\theta}| \geq |\vev{\theta}|_{DSW} = \xi$, and $\xi \sim (0.1-0.01)\, M_{Pl}$. For coefficients $c_1$, $c_2$ and $c_3$ of order one, this solution gives $\vev{\chi_i} / M_{Pl}$ of order one, which is unacceptable in the context of an effective field theory. In addition, when higher order terms are added in (\ref{eq:W_singlets}), the vacuum (\ref{eq:vacuum_singlets}) changes.

\section{Supersymmetry breaking and low-energy vacuum}
\label{section:susy}
\indent

The purpose of this section is to discuss how supersymmetry breaking affects the conclusions from the previous section. Since models with an anomalous $U(1)$ have numerous implications for low-energy phenomenology, it is indeed essential to ensure that the flat direction analysis is relevant to the determination of the low-energy vacuum.

The scalar potential of the low-energy theory reads:
\begin{equation}
  V\ =\ \frac{1}{2}\, \sum_a\ D^a D^a\ +\ \sum_i\ \svev{F_i}\ +\ V_{\!\!\! \scriptscriptstyle{\barre{susy}}}
\end{equation}
Since supersymmetry has to be broken in a soft way, $V_{\!\!\! \scriptscriptstyle{\barre{susy}}}\ $ is of the form $\widetilde{m}\, V^{(3)}\, +\, \widetilde{m}^2\, V^{(2)}$, where $\widetilde{m} \sim 1\, TeV$ is the scale of supersymmetry breaking, and $V^{(2)}$, $V^{(3)}$ are functions of the scalar fields with dimensions 2 and 3, respectively\footnote{This definition allows for higher order terms suppressed by negative powers of the Planck mass, e.g. $\Phi^n / M^{n-2}_{Pl} \in V^{(2)}$.}. This has the obvious consequence that the actual minimum of the scalar potential is close to a flat direction of the supersymmetric theory; otherwise the $D$-terms and $F$-terms would give a positive contribution of order $\xi^4$ to $V$, while $V_{\!\!\! \scriptscriptstyle{\barre{susy}}}\ $ would contribute at most as $\widetilde{m}\, \xi^3$, with no possibility of compensation. We thus necessarily have $\vev{F_i} \ll \xi^2$ and $\vev{D^a} \ll \xi^2$, which implies in particular that the relations $\svev{\theta_{\alpha}} \geq \svev{\theta_{\alpha}}_{DSW}$ still hold (provided that the necessary conditions are satisfied), and that fields charged under $G$ with vevs of order $\xi$ should be aligned in the sense of Eq. (\ref{eq:alignment2}). In addition, we shall assume that there are no compensations inside the $F$-terms, which means that all contributions to the $F$-terms must be much smaller than $\xi^2$.

Let us first consider the flat directions for which $\svev{\theta_{\alpha}} \geq \svev{\theta_{\alpha}}_{DSW}$ hold for all $\alpha$. The minimization procedure amounts to adjusting the field vevs around this flat direction so as to obtain the lowest value for the scalar potential; as a result, some fields acquire vevs of the order of the supersymmetry breaking scale $\widetilde{m}$, or of an intermediate scale such as $(\widetilde{m}\, M_{Pl})^{1/2}$. Clearly those cannot be the $\theta$ fields. In addition, the $\chi$ and $\Phi$ fields cannot have a vev of order $\xi$, otherwise some invariant of the form $S\, \theta^{n_0}_0 \theta^{n_1}_1 \ldots \theta^{n_N}_N$ or $\chi^m\, \theta^{m_0}_0 \theta^{m_1}_1 \ldots \theta^{m_N}_N$ would give a contribution of order $\xi^2$ to the $F$-terms. As a result, the low-energy vacuum is a slight deviation from the DSW vacuum:
\begin{equation}
  \svev{\theta_{\alpha}}\ \simeq\ \svev{\theta_{\alpha}}_{DSW}\  \hskip 1.5cm  \vev{\Phi_i}\, ,\ \vev{\chi_i}\ \ll\ \xi^2
\label{eq:DSW_low}
\end{equation}
This is perfectly consistent with the conclusions from the flat direction analysis: flat directions along which $\svev{\theta_{\alpha}} \geq \svev{\theta_{\alpha}}_{DSW}$ for all $\alpha$ are lifted down to the DSW vacuum by the $F$-terms, and the only effect of supersymmetry breaking is to give a small vev to the $\chi$ and $\Phi$ fields. Note that the symmetries of $G$ are not broken at the scale $\xi$.

Consider now the flat directions for which $\svev{\theta_{\alpha}} \geq \svev{\theta_{\alpha}}_{DSW}$ does not hold for all $\alpha$. Contrary to the previous type of flat directions, these may be only partially lifted at the supersymmetric level, and the effect of supersymmetry breaking is to lift them completely, leading possibly to undesired vacua, as we illustrate bellow. For this purpose, we go back to the example of Subsection \ref{subsubsection:single_S} and consider the flat direction $\vev{Q_1, \bar u_1, H_u, \theta_0, \theta_1}$ associated with the non-holomorphic invariant $Q_1 \bar u_1 H_u\, \theta^{-1}_0 \theta^{3}_1$. Since along this flat direction $\svev{\theta_{0}} < \svev{\theta_{0}}_{DSW}$, we can see from (\ref{eq:F_u2}) that, in the case $n_0>0$, $F_{\bar u_2}$ can be small compared to $\xi^2$ while $v^2_S \equiv \svev{Q_1} = \svev{\bar u_1} = \svev{H_u}$ is of order $\xi^2$. For example, in the case $n_0=1$, the minimization of the scalar potential could yield $\vev{\theta_0} \sim \widetilde{m}$ (with $\svev{F_{\bar u_2}} \sim \widetilde{m}^2\, \xi^2$ being compensated for by e.g. $\widetilde{m}^2_{H_u}\, \svev{H_u}$, with $\widetilde{m}^2_{H_u}$ negative). In this case, the Standard Model symmetries would be broken at the scale $\xi$. We conclude that such flat directions are potentially dangerous and should be lifted at the supersymmetric level, in the way that has been discussed in the previous section. For instance, the problem disappears in the $n_0=0$ case, since $v_S \sim \xi$ would give $F_{\bar u_2} \sim \xi^2$, which as stressed before cannot be the case at the minimum of the scalar potential.

From this qualitative discussion, we conclude that supersymmetry breaking does not change radically the conclusions from the flat direction analysis. Its two main effects are to modify slightly the DSW vacuum by giving small or intermediate vevs to the $\chi$ and $\Phi$ fields, and to lift the flat directions that are present in the supersymmetric theory. It is therefore essential to check, in a specific model, that the flat directions that may lead to undesired vacua are completely lifted already in the supersymmetric limit.

\section{Summary of the procedure}
\label{section:procedure}
\indent

We now summarize the generic procedure to analyze flat directions in anomalous $U(1)$ models that satisfy conditions (a) and (b):
\begin{enumerate}
  \item find a basis $S_1$, \ldots, $S_p$ for holomorphic $G$-invariants involving only $\Phi$ fields; add to this basis all $G$-singlets $\chi_1$, \ldots, $\chi_q$.
  \item for each element of the basis, compute the set of numbers ($n_0$, \ldots, $n_N$) defined by (\ref{eq:condition2}). The corresponding $D$-flat direction is $\vev{S\,; \vec{\theta}\,}$ in the case of a $G$-invariant, and $\vev{\chi\,; \vec{\theta}\,}$ in the case of a $G$-singlet; the vevs of the $\theta$ fields are determined by (\ref{eq:Dflat}), and they satisfy the constraints:
\begin{equation}
  \left\{  \begin{array}{lll}
  \svev{\theta_{\alpha}}\ \geq\ \svev{\theta_{\alpha}}_{DSW} & \hskip 1cm & n_{\alpha}\ \geq\ 0  \\
    &  &  \\
  \svev{\theta_{\alpha}}\ <\ \svev{\theta_{\alpha}}_{DSW} & \hskip 1cm & n_{\alpha}\ <\ 0
  \end{array}  \right.
\end{equation}
  \item the most general $D$-flat direction involving a set of fields (\{$\Phi_i$\}; \{$\chi_i$\}) is parameterized by the vevs of the $\chi_i$ and the $\Phi_i$, the latter being subject to constraints (\ref{eq:alignment2}). The vevs of the $\theta$ fields are determined by (\ref{eq:Dflat2}), which implies
\begin{equation}
\svev{\theta_{\alpha}}\ \geq\ \svev{\theta_{\alpha}}_{DSW} \hskip 1cm \mbox{if} \hskip 1cm  n^a_{\alpha},\ n^i_{\alpha}\ \geq\ 0
\label{eq:condition}
\end{equation}
for all relevant $S_a$ and $\chi_i$.
  \item determine which flat directions are lifted by $F$-terms. Two cases must be distinguished:

- flat directions for which all relevant \{$n^a_{\alpha}\,$; $n^i_{\alpha}$\} are positive are lifted down to the $DSW$ vacuum by $F$-terms.

- other flat directions are only partially lifted, unless invariants of the form $S'\, \theta^{n'_0}_0 \theta^{n'_1}_1 \ldots \theta^{n'_N}_N$, where $S'$ contains no more than one field that does not have a vev, are present in the superpotential. For this to happen, the $n'_{\alpha}$ must satisfy the conditions specified in Subsection \ref{subsubsection:generic}. Several such invariants are in general necessary to lift completely the flat direction.
  \item once the presence of the invariants required to lift a given flat direction has been checked, a more carefull analysis should take into account the possibility of cancellations inside the $F$-terms, and show that the flat direction is indeed lifted at some order. Also, one should check that cancellations do not allow other stable vacua of singlets than the DSW solution, even though such a possibility seems to be very unlikely.
\end{enumerate}

\section{The vacuum of a realistic low-energy model}
\label{section:model}
\indent

In the top-bottom approach of constructing models with an anomalous $U(1)$, one 
typically obtains upon
compactification of a superstring a certain set of fields that here we called 
``$G$-singlet''.
Then this set has to be divided into a subset that takes a vev at a scale $~\xi$ 
(``$\theta$ set'') and a subset that
does not (``$\chi$ set''). Unfortunately, in realistic model building this 
separation has turned out to be 
far from unique. From the vacuum point of view therefore 
the minimalistic, bottom-up approach is advantageous, in the sense that one can
ensure that the $\theta$-set is unique. The simplest way to achieve this in a 
model with an anomalous, flavor blind 
$U(1)_X$, is to allow vevs only to those $G$-singlets that have nonnegative X 
charges. 
A model with such property has been constructed \cite{E6IR}. Here, we will look 
only at its vacuum
structure by applying the systematic analysis of section 
\ref{section:procedure}. 
The gauge group of the model is $G = SU(3)_c\times SU(2)_w\times U(1)_Y$
augmented by 
three additional $U(1)$'s, one of which is anomalous and flavor blind ($X$)
and the other two ($Y_1$ and $Y_2$) are nonanomalous, flavor dependent, 
combinations
of the generators $Y$, $V$ and $V'$ of $E_6$. 
The field content is three families of fields in the ${\bf 27}$ of $E_6$ plus 
several $G$-singlets, among which three
$\theta_{\alpha}$ fields with charges that form the matrix
\begin{equation}
A= \left( \begin{array}{ccc}
1&0&0\\ 0&-1&1\\ 1&-1&0
\end{array}  \right)\ 
\end{equation} 
where the first row is their X-charges, the second and the third rows are their 
$Y_1$ and $Y_2$ charges respectively.  
The fact that $\det A \ne 0$ and that the first 
column of $A^{-1}$ is $(1,1,1)$, ensures the existence of the
vacuum configuration $\vev{\theta_1,\theta_2,\theta_3}$.
Then, we have to extend the basis of MSSM holomorphic invariants of Ref. \cite{Martin} to take into account the presence of vector-like matter in the model, and compute the powers of the $\theta$ fields for each basis invariant, as well as for each Standard Model singlet. In addition we must check that the superpotential does not contain a term linear 
in a $\chi$ field
(such a term would be 
for example ${\overline N}_i\theta_1^{n_1}\theta_2^{n_2}\theta_3^{n_3}$ with 
$n_1, n_2, n_3$ nonnegative integers).  
It turns out that, for a convenient choice of the $X$-charge, all the supersymmetric zeroes of this model reside in the sector 
that contains the fields lying in the ${\bf 16}$ of $SO(10)$.   
A list of all the supersymmetric zeroes of the model can be found in table
1. It is remarkable that among the total number of invariants which is of order 
of thousands we find only a
few zeroes due to negative powers. The first column of the table contains the 
basis invariant. 
In the second column, by solving (\ref{eq:condition2}), we compute the powers of the $\theta $ 
fields corresponding to that basis invariant.  
The third column shows the potentially dangerous flat direction associated with the invariant
of the first column, and
the fourth column an holomorphic invariant that lifts it. 
\par Finally we would like to stress the fact that in this example we checked 
flat directions that
correspond only to basis invariants. However, as mentioned in the general 
discussion, in a
complete analysis one has to look at all possible invariant polynomials 
containing negative powers
and their associated ``composite'' flat directions,
and check that the invariants present in the superpotential actually lift them. Such an analysis is beyond the scope of this paper.

\section{Conclusion}
\label{section:conclusion}
\indent

In this letter, we exposed a systematic procedure to study flat directions in a large class of anomalous $U(1)$ models. Those models, whose gauge group contains a set of Abelian horizontal symmetries connecting the observable sector to the hidden sector, are prototypes of models of mass hierarchies. We stated the conditions ensuring the uniqueness of the DSW vacuum that breaks those symmetries, so that the model gives definite predictions for mass hierarchies. More generally, we gave a very simple way to identify flat directions that may lead to other vacua, and studied under which conditions they are lifted by $F$-terms. We also checked that supersymmetry breaking only slightly modifies the conclusions from the flat direction analysis. In particular, when all potentially dangerous flat directions are already lifted at the supersymmetric level, the only expected effect of supersymmetry breaking is to shift slightly the DSW vacuum by giving small or intermediate vevs to fields that are not involved in it.

The structure of the models we considered is characteristic of a large class of effective string theories. In principle, one can apply our procedure to these (provided that conditions (a) and (b) are fulfilled), but some of our conclusions may be escaped, because discrete symmetries and string selection rules prevent a lot of holomorphic invariants from appearing in the superpotential. In particular, the uniqueness of the DSW vacuum - which is of great phenomenological significance - is not ensured in generic string models.

\vskip 1cm 
{\bf Acknowledgements}
\vskip .5cm
After this work was completed, a related paper \cite{Cleaver} appeared, which presented general techniques to classify flat directions of non-Abelian singlets in a large class of heterotic superstring models with an anomalous $U(1)$ .

We wish to thank Pierre Ramond for very helpfull suggestions and comments. S.L. thanks the Institute for Fundamental Theory, Gainesville, for its hospitality and financial support.



\vfill
\eject
\newpage


\begin{table}
\vskip -2cm
\caption{{\bf Flat Directions of Model of Ref. \cite{E6IR}}}
\vskip 1cm
\hskip -2.2cm
\begin{tabular}{|c|c|c|c|}

           \hline  & & &  \\
$~{\rm {\bf Basis ~Invariant}}~$ & $ {\bf (n_1,n_2,n_3)} $ & $~{\bf Flat~ Direction}~$ 
& 
$~{\bf Invariant~in~W~ that~ lifts~ the ~FD}~$\\  & & &  \\
                       \hline \hline   & & &  \\
$~{\overline N}_3$&$ (3/2,-1/2,1/2) ~$ & 
$~<{\overline N}_3,\theta_1,\theta_3>~$ & 
${\overline N}_2{\overline N}_3^{\, 3}\, \theta_1^{6}\theta_3^{4}~$ \\
                                      & & &  \\ 
                                                                 
${\bf L}_1{\bf H_u}$&$ (-3/2,1/2,5/2) ~$ & 
$~<{\bf L}_1,{\bf H_u},\theta_2,\theta_3>~$ & 
$~{\bf L}_1{\overline N}_3{\bf H_u}\theta_3^{3}~$\\ 
                                      & & &  \\
${\bf L}_2{\bf H_u}$&$ (-3/2,1/2,-1/2)~$ & 
$~<{\bf L}_2,{\bf H_u},\theta_1 (\theta_3),\theta_2>~$ & 
$~{\bf L}_2{\overline N}_3{\bf H_u}~$\\           
                                                             & & &  \\  
${\bf L}_3{\bf H_u}$&$ (-3/2,1/2,-1/2)~$ & 
$~<{\bf L}_3,{\bf H_u},\theta_1 (\theta_3),\theta_2>~$ & 
$~{\bf L}_3{\overline N}_3{\bf H_u}~$\\ 

                                                        & & &  \\  

${\bf L}_2{\overline 
e}_1{\bf H_d}$&$ (3,2,-1)~$ &
$<{\bf L}_2,{\overline e}_1,{\bf H_d},\theta_1,\theta_2>~$ &
${\bf L}_2{\overline e}_3{\bf H_d}\theta_1^{3}$\\
 
                                                                 & & &  \\

${\bf L}_3{\overline 
e}_1{\bf H_d}$&$ (3,2,-1)~$ &
$<{\bf L}_3,{\overline e}_1,{\bf H_d},\theta_1,\theta_2>~$ &
${\bf L}_3{\overline e}_3{\bf H_d}\theta_1^{3}$\\
 
                                                                 & & &  \\

${\bf L}_2{\bf L}_3{\overline 
e}_1$&$ (3/2,5/2,-3/2)~$ &
$<{\bf L}_2,{\bf L}_3,{\overline e}_1,\theta_1,\theta_2>~$ &
${\bf L}_2{\bf L}_3{\overline e}_1{\overline N}_1\theta_1^{3}\theta_2^{6}$\\
 
                                                                 & & &  \\
${\bf L}_2{\bf L}_3{\overline 
e}_3$&$ (3/2,1/2,-1/2)~$ &
$<{\bf L}_2,{\bf L}_3,{\overline e}_3,\theta_1,\theta_2>~$ &
${\bf L}_2{\bf L}_3{\overline e}_3{\overline N}_3\theta_1^{3}$\\  
 
                                                     & & &  \\

${\bf L}_2{\bf Q}_3{\bf {\overline 
d}_2}$&$ (3/2,1/2,-1/2)~$ &
$<{\bf L}_2,{\bf Q}_3,{\bf {\overline d}_2},\theta_1,\theta_2>~$ &
${\bf L}_2{\bf Q}_3{\bf {\overline d}_2}{\overline N}_3
\theta_1^{3}$\\ 
                                                                & & &  \\  
${\bf L}_3{\bf Q}_3{\bf {\overline 
d}_2}$&$ (3/2,1/2,-1/2)~$ &
$<{\bf L}_3,{\bf Q}_3,{\bf {\overline d}_2},\theta_1,\theta_2>~$ &
${\bf L}_3{\bf Q}_3{\bf {\overline d}_2}{\overline N}_3
\theta_1^{3}$\\ 
 
                                                                 & & &  \\  
${\bf L}_2{\bf Q}_3{\bf {\overline 
d}}_3$&$ (3/2,1/2,-1/2)~$ &
$<{\bf L}_2,{\bf Q}_3,{\bf {\overline d}}_3,\theta_1,\theta_2>~$ &
${\bf L}_2{\bf Q}_3{\bf {\overline d}}_3{\overline N}_3\theta_1^{3}$\\ 
                                                                & & &  \\
${\bf L}_3{\bf Q}_3{\bf {\overline 
d}}_3$&$ (3/2,1/2,-1/2)~$ &
$<{\bf L}_3,{\bf Q}_3,{\bf {\overline d}}_3,\theta_1,\theta_2>~$ &
${\bf L}_3{\bf Q}_3{\bf {\overline d}}_3{\overline N}_3\theta_1^{3}$\\  
                                                                & & &  \\
${\bf {\overline u}}_3{\bf {\overline d}}_2
{\bf {\overline d}}_3$&$ (3/2,1/2,-1/2)~$ &
$<{\bf {\overline u}}_3,{\bf {\overline d}}_2,
{\bf {\overline d}}_3,\theta_1,\theta_2>~$ 
&
${\bf {\overline u}}_3{\bf {\overline d}}_2{\bf {\overline d}}_3
{\overline N}_3\theta_1^{3}$\\ 
                                                                & & &  \\  
   
${\bf Q}_3{\bf {\overline u}}_3{\overline e}_1{\bf 
H_d}$&$ (9/2,3/2,-1/2)~$ &
$<{\bf Q}_3,{\bf {\overline u}}_3,{\overline e}_1,{\bf H_d},\theta_1,\theta_2>~$ 
&
${\bf Q}_3{\bf {\overline u}}_3{\bf H_u}$\\ 
                                                                & & &  \\  
${\bf Q}_3{\bf {\overline u}}_3{\overline e}_3{\bf 
H_d}$&$ (9/2,-1/2,1/2)~$ &
$<{\bf Q}_3,{\bf {\overline u}}_3,{\overline e}_3,{\bf H_d},\theta_1,\theta_3>~$ 
&
${\bf Q}_3{\bf {\overline u}}_3{\bf H_u}$\\
                                                                & & &  \\   
${\bf Q}_3{\bf {\overline u}}_3{\bf L}_2{\overline 
e}_1$&$ (3,2,-1)~$ &
$<{\bf Q}_3,{\bf {\overline u}}_3,{\bf L}_2,{\overline e}_1,\theta_1,\theta_2>~$ 
&
${\bf Q}_3{\bf {\overline u}}_3{\bf H_u}$\\
                                                                & & &  \\  
${\bf Q}_3{\bf {\overline u}}_3{\bf L}_3{\overline 
e}_1$&$ (3,2,-1)~$ &
$<{\bf Q}_3,{\bf {\overline u}}_3,{\bf L}_3,{\overline e}_1,\theta_1,\theta_2>~$ 
&
${\bf Q}_3{\bf {\overline u}}_3{\bf H_u}$\\
                                                                & & &  \\ 
${\bf Q}_3{\bf {\overline u}}_3{\bf Q}_3{\bf {\overline u}}_3{\overline 
e}_1$&$ (9/2,3/2,-1/2)~$ &
$<{\bf Q}_3,{\bf {\overline u}}_3,{\overline e}_1,\theta_1,\theta_2>~$ &
${\bf Q}_3{\bf {\overline u}}_3{\bf H_u}$\\
                                                                & & &  \\  
${\bf Q}_3{\bf {\overline u}}_3{\bf Q}_3{\bf {\overline u}}_3{\overline 
e}_3$&$ (9/2,-1/2,1/2)~$ &
$<{\bf Q}_3,{\bf {\overline u}}_3,{\overline e}_3,\theta_1,\theta_3>~$ &
${\bf Q}_3{\bf {\overline u}}_3{\bf H_u}$\\
                                                                 & & &  \\   
${\bf {\overline d}}_1{\bf {\overline d}}_2{\bf {\overline d}}_3{\bf L}_2{\bf 
L}_3
$&$ (3/2,3/2,-1/2)~$ &
$<{\bf {\overline d}}_1,{\bf {\overline d}}_2,{\bf {\overline d}}_3,
{\bf L}_2,{\bf L}_3,\theta_1,\theta_2>~$ &
${\bf {\overline d}}_1{\bf {\overline d}}_2{\bf {\overline d}}_3
{\bf L}_2{\bf L}_3{\overline N}_3 \theta_1^{3}\theta_2~$\\  
 
                                                      & & &  \\     \hline

\end{tabular}
\end{table}



\begin{thebibliography}{Ref}

\bibitem{string_models} T. Kobayashi and H. Nakano, Nucl. Phys. B496 (1997) 103; A. Faraggi and G. Cleaver, hep-ph/9711339. See also Ref. \cite{Faraggi} to \cite{Tye}.

\bibitem{Faraggi} A. Faraggi, Nucl. Phys. B387 (1992) 239, Nucl. Phys. B403 (1993) 101, Nucl. Phys. B407 (1993) 57; A.E. Faraggi and E. Halyo, Nucl. Phys. B416 (1994) 63.

\bibitem{Kim} H. B. Kim and J. E. Kim, Phys. Lett. B300 (1993) 343.

\bibitem{Chaudhuri} S. Chaudhuri, G. Hockney and J. Lykken, Nucl. Phys. B496 (1996) 357.

\bibitem{Tye} Z. Kakushadze and S.-H. H. Tye, Phys. Lett. B392 (1997) 335, Phys. Rev. D55 (1997) 7896.

\bibitem{fermion_masses} L. Ib\'a\~nez and G. G. Ross, Phys. Lett. B332 (1994) 100; P. Bin\'etruy and P. Ramond, Phys. Lett. B350 (1995) 49; Y. Nir, Phys. Lett. B354 (1995) 107; V. Jain and R. Shrock, Phys. Lett. B352 (1995) 83; E. Dudas, S. Pokorski and C.A. Savoy, Phys. Lett. B356 (1995) 45; P. Bin\'etruy, S. Lavignac,  and P. Ramond,  Nucl. Phys. B477 (1996) 353; P. Ramond, in ``Frontiers in Quantum Field Theory'', H. Itoyama, M. Kaku, H. Kunimoto, N. Ninomya \& H. Shirokura, eds, (World Scientific, 1996).

\bibitem{susy_breaking} P. Bin\'etruy and E. Dudas, Phys. Lett. B389 (1996) 503; G. Dvali and A. Pomarol, Phys. Rev. Lett. 77 (1996) 3728; R.N. Mohapatra and A. Riotto, Phys. Rev. D55 (1997) 1138; {\it ibid.}, 4262.

\bibitem{BILR} P. Bin\'etruy, N. Irges, S. Lavignac and P. Ramond, Phys. Lett. B403 (1997) 38.

\bibitem{inflation} P. Bin\'etruy and G. Dvali, Phys. Lett. B388 (1996) 241; E. Halyo, Phys. Lett. B387 (1996) 43.

\bibitem{Green-Schwarz} M. Green and J. Schwarz, Phys. Lett. B149 (1984) 117.

\bibitem{Atick} J. Atick, L. Dixon and A. Sen, Nucl. Phys. B292 (1987) 109.

\bibitem{DSW} M. Dine, N. Seiberg and E. Witten, Nucl. Phys. B289 (1987) 317.
 
\bibitem{DGPS} E. Dudas, C. Grojean, S. Pokorski and C.A. Savoy, Nucl. Phys. B481 (1996) 85.

\bibitem{Froggatt-Nielsen} C. Froggatt and H.B. Nielsen Nucl. Phys. B147 (1979) 277; M. Leurer, Y. Nir, and N. Seiberg, Nucl. Phys. B398 (1993) 319.

\bibitem{neutrinos}  H. Dreiner, G.K. Leontaris, S. Lola, G.G. Ross and C. Scheich, Nucl. Phys. B436 (1995) 461; Y. Grossman and Y. Nir, Nucl. Phys. B448 (1995) 30; P. Bin\'etruy, S. Lavignac, S. Petcov and P. Ramond, Nucl. Phys. B496 (1997) 3.

\bibitem{FCNC} M. Leurer, Y. Nir, and N. Seiberg, Nucl. Phys. B420 (1994) 468; E. Dudas, S. Pokorski and C.A. Savoy, Phys. Lett. B369 (1996) 255; Y. Kawamura and T. Kobayashi, Phys. Lett. B375 (1996) 141; Y. Kawamura, T. Kobayashi and T. Komatsu, Phys. Lett. B400 (1997) 284; A.E. Nelson and D. Wright, Phys. Rev. D56 (1997) 1598.

\bibitem{B_and_L}  V. Ben-Hamo and Y. Nir, Phys. Lett. B339 (1994) 77; T. Banks, Y. Grossman, E. Nardi and Y. Nir, Phys. Rev. D52 (1995) 5319; F. Borzumati, Y. Grossman, E. Nardi and Y. Nir, Phys. Lett. B384 (1996) 123; P. Bin\'etruy, E. Dudas, S. Lavignac and C. Savoy, hep-ph/9711517.

\bibitem{EIR} J.K. Elwood, N. Irges and P. Ramond, hep-ph/9705270.

\bibitem{Buccella} F. Buccella, J.-P. Derendinger, S. Ferrara and C.A. Savoy, Phys. Lett. B115 (1982) 375.

\bibitem{Gatto} C. Procesi and G.W. Schwarz, preprint Department of Mathematics, Brandeis University (1985); see also: R. Gatto and G. Sartori, Phys. Lett. B157 (1985) 389.

\bibitem{Martin} T. Gherghetta, C. Kolda and S.P. Martin, Nucl. Phys. B468 (1996) 37.

\bibitem{E6IR} N. Irges, S. Lavignac and P. Ramond, in preparation.

\bibitem{Cleaver} G. Cleaver, M. Cveti\v{c}, J.R. Espinosa, L. Everett and P. Langacker, hep-th/9711178.


\end{thebibliography}
\end{document}